\documentstyle[12pt]{article}
\newcounter{eq}
\newcounter{sc}

\def\overleftrightarrow#1{\vbox{\ialign{##\crcr
 $\leftrightarrow$\crcr\noalign{\kern-1pt\nointerlineskip}
 $\hfil\displaystyle{#1}\hfil$\crcr}}}

\newlength{\minitwocolumn}
\begin{document}

%%%%%%%%%%%%%%%%%%%%%%%%%%%%%%%%%%%%%%%%%%%%%%%%%%%%%%%%%%%%%%%%%%
%%%%%%%%%%%%%%%%%%%%%%%% Title %%%%%%%%%%%%%%%%%%%%%%%%%%%%%%%%%%%
%%%%%%%%%%%%%%%%%%%%%%%%%%%%%%%%%%%%%%%%%%%%%%%%%%%%%%%%%%%%%%%%%%
\begin{flushright}
DPUR/TH/41\\
July, 2014\\
\end{flushright}
\vspace{20pt}

%\magnification=\magstep1
\pagestyle{empty}
\baselineskip15pt
%\font\cmssB=cmss17
%\font\cmssS=cmss10

\begin{center}
{\large\bf Conformal Higgs Inflation
\vskip 1mm }

\vspace{20mm}
Ichiro Oda \footnote{E-mail address:\ ioda@phys.u-ryukyu.ac.jp}
and Takahiko Tomoyose \footnote{E-mail address:\ k138335@eve.u-ryukyu.ac.jp}

\vspace{5mm}
           Department of Physics, Faculty of Science, University of the 
           Ryukyus,\\
           Nishihara, Okinawa 903-0213, Japan.\\

\end{center}

%\maketitle

\vspace{5mm}
\begin{abstract}
We investigate a locally scale-invariant (that is, Weyl-invariant) theory which describes the coupling 
of gravity and the standard model from the viewpont of the Higgs mechanism and inflation.
It is shown that this theory exhibits a peculiar feature of a coupling between the gauge field and
the inflaton in a large field limit whereas it nicely describes the standard model coupled to
general relativity in a small field limit. Moreover,  we discuss a possibility that this Weyl invariant 
theory could explain the recent BICEP2 measurement by deforming the potential term in an appropriate way, 
and find it to be difficult to do that even if we introduce a non-analytical type of the potential.   
\end{abstract}

\newpage
\pagestyle{plain}
\pagenumbering{arabic}
%\setcounter{page}{1}

%%%%%%%%%%%%%%%%%%%%%%%%%%%%%%%%%%%%%%%%%%%%%%%%%%%%%%%%%%%%%%%%%%
%%%%%%%%%%%%%%%%%%%%%%%% Article %%%%%%%%%%%%%%%%%%%%%%%%%%%%%%%%%
%%%%%%%%%%%%%%%%%%%%%%%%%%%%%%%%%%%%%%%%%%%%%%%%%%%%%%%%%%%%%%%%%%

\rm
%%%%%%%%%%%%%%%%%%%%%%%%%%%%%%%%%%%%%%%%%%%%%%%%%%%%%%%%%%%%%%%%%%%%%
%%%%%%%%%%%%%%%%%%%%%%%%%%%%%%   SEC  1    %%%%%%%%%%%%%%%%%%%%%%%%%%
%%%%%%%%%%%%%%%%%%%%%%%%%%%%%%%%%%%%%%%%%%%%%%%%%%%%%%%%%%%%%%%%%%%%%
\section{Introduction}

In attempting to construct a theory beyond the standard model and a theory of inflation \cite{Guth, Linde1}, 
it is desirable for us to be guided by some fundamental principles and symmetries such as 
the equivalence principle and the general coordinate invariance in general relativity. 
As one of such fundamental symmetries, we often encounter a scale or conformal symmetry 
in particle physics and cosmology. 

For instance, on the microscopic scale of particle physics, it is well known that the standard model 
has a global scale symmetry if the (negative) Higgs mass term is neglected. This fact might suggest that
the underlying symmetry behind the standard model would be a global or local scale symmetry and the Higgs 
mass term could emerge as the result of the spontaneous symmetry breakdown of the scale symmetry \cite{Bardeen}. 
In addition, even in cosmology, the scale symmetry plays a critical role. For instance, the cosmic mocrowave background 
radiation has almost scale-invariant fluctuations \cite{Planck} so it
seems to be natural at least for the present authors to conjecture that the universe was
controlled by the scale symmetry around the beginning of its creation. Together with these two
examples from particle physics and cosmology, it seems to be of interest to pursue an idea such that a theory 
beyond the standard model is not only scale-invariant but also couples to a scale-invariant gravity 
in order to account for inflation scenario.

The inflaton is a unknown scalar field which triggers the inflationary expansion and must be very weakly
interacting in order to explain the observed isotropy of the microwave background radiation. A natural question
to be asked frequently is whether the inflaton could be identified with recently discovered Higgs particle \cite{ATLAS, CMS}.
The recent revival of interest to this idea, dubbed Higgs inflation, is based on the observation that 
when the simple chaotic inflation with the potential $\lambda \phi^4$ is combined with the non-minimal 
gravitational coupling $\xi \phi^2 R$ in the Jordan frame, the inflation model can explain the observational data 
if $\lambda = {\cal O}(1)$ and $\xi \approx 10^5$ \cite{Bezrukov}.
 
The non-minimal gravitational coupling $\xi \phi^2 R$ is invariant under a global scale transformation \cite{Oda1, Oda2, Oda3}, 
so there might exist a possibility of generalizing the Higgs inflation to a locally scale-invariant gravitational
theory. Indeed, in recent years, a locally scale-invariant class of inflation models has been established and
discussed from various viewpoints in \cite{Kallosh1}-\cite{Costa}, but a detailed analysis has not yet done for
the standard model embeded in a locally scale-invariant gravity.   
In this article, we construct a theory of the standard model coupled to the Weyl-invariant gravity. In particular,
we discuss the Higgs mechanism in this theory and explore the question of whether this theory could explain the recent remarkable 
results of the BICEP2 \cite{BICEP2}. 

The structure of this article is the following: In Section 2, we present a simple model which accomodates
a local scale symmetry and the abelian gauge symmetry and explain the formalism in the Einstein
gauge and the Jordan gauge. In Section 3, we comment on an $SO(1,1)$ accidental symmetry and a shift symmetry.
In Section 4, we discuss the Higgs mechanism and derive the value of mass of the gauge field. 
In Section 5, the abelian gauge model constructed in Section 2 is generalized to a model with the non-abelian 
gauge field. 
In Section 6, we argue inflation derived from the abelian gauge model and show that it is difficult to account
for the BICEP2 results. 
We conclude in Section 7.

%%%%%%%%%%%%%%%%%%%%%%%%%%%%%%%%%%%%%%%%%%%%%%%%%%%%%%%%%%%%%%%%%%%%%
%%%%%%%%%%%%%%%%%%%%%%%%%%%%%%   SEC  2    %%%%%%%%%%%%%%%%%%%%%%%%%%
%%%%%%%%%%%%%%%%%%%%%%%%%%%%%%%%%%%%%%%%%%%%%%%%%%%%%%%%%%%%%%%%%%%%%
\section{An abelian gauge model}

Let us start with an abelian model where a  $U(1)$ gauge field $A_\mu$ couples to a Weyl-invariant
gravity of two scalar fields, one of which, $\phi$, is real while the other, $H$ which is nothing but the
Higgs doublet field, is complex. In this article, we will ignore fermions completely since their existence 
does not change our conclusion at all. 

With a background curved metric $g_{\mu\nu}$, the Lagrangian takes the form\footnote{We follow 
notation and conventions by Misner et al.'s textbook \cite{MTW}, for instance, the flat Minkowski metric
$\eta_{\mu\nu} = diag(-, +, +, +)$, the Riemann curvature tensor $R^\mu \ _{\nu\alpha\beta} = 
\partial_\alpha \Gamma^\mu_{\nu\beta} - \partial_\beta \Gamma^\mu_{\nu\alpha} + \Gamma^\mu_{\sigma\alpha} 
\Gamma^\sigma_{\nu\beta} - \Gamma^\mu_{\sigma\beta} \Gamma^\sigma_{\nu\alpha}$, 
and the Ricci tensor $R_{\mu\nu} = R^\alpha \ _{\mu\alpha\nu}$.
The reduced Planck mass is defined as $M_p = \sqrt{\frac{c \hbar}{8 \pi G}} = 2.4 \times 10^{18} GeV$.
Through this article, we adopt the reduced Planck units where we set $c = \hbar = M_p = 1$. 
In this units, all quantities become dimensionless. 
Finally, note that in the reduced Planck units, the Einstein-Hilbert Lagrangian density takes the form
${\cal L}_{EH} = \frac{1}{2} \sqrt{-g} R$.}:
%**   Lagr 1  %%%%%%%%%%%%%%%%%%%%%%%%%%%%%%%%%%%%%%%%%%%%%%%%%%%%%%%%%
\begin{eqnarray}
\frac{1}{\sqrt{-g}}{\cal L} &=&  \frac{1}{12} \left( \phi^2 - 2 |H|^2 \right) R + \frac{1}{2} g^{\mu\nu} 
\partial_\mu \phi \partial_\nu \phi - g^{\mu\nu} (D_\mu H)^\dagger (D_\nu H)    \nonumber\\
&-& \frac{1}{4} g^{\mu\nu} g^{\rho\sigma} F_{\mu\rho} F_{\nu\sigma} - V(\phi, H),
\label{Lagr 1}
\end{eqnarray}
%%%%%%%%%%%%%%%%%%%%%%%%%%%%%%%%%%%%%%%%%%%%%%%%%%%%%%%%%%%%%%%%%%%
where $\phi$ is a ghost, but this is not a problem because it can be removed by fixing the Weyl symmetry. 
The covariant derivative and field strength in the abelian gauge group are respectively defined as 
%**   Def 1  %%%%%%%%%%%%%%%%%%%%%%%%%%%%%%%%%%%%%%%%%%%%%%%%%%%%%%%%%
\begin{eqnarray}
D_\mu H = (\partial_\mu + i g A_\mu) H,   \quad
(D_\mu H)^\dagger = H^\dagger (\overleftarrow{\partial}_\mu - i g A_\mu) ,   \quad
F_{\mu\nu} = \partial_\mu A_\nu - \partial_\nu A_\mu,
\label{Def 1}
\end{eqnarray}
%%%%%%%%%%%%%%%%%%%%%%%%%%%%%%%%%%%%%%%%%%%%%%%%%%%%%%%%%%%%%%%%%%%
with $g$ being a $U(1)$ real coupling constant. (We use the same alphabet "$g$" to denote the gauge coupling
and the determinant of the metric tensor, but the difference would be obvious since the latter always appears
in the form of $\sqrt{-g}$.)

The Lagrangian (\ref{Lagr 1}) is invariant under a local scale transformation (or Weyl transformation). In fact, 
with a local parameter $\Omega(x)$ the Weyl transformation is defined as 
%**   Weyl transf %%%%%%%%%%%%%%%%%%%%%%%%%%%%%%%%%%%%%%%%%%%%%%%%%%%%%%%%%
\begin{eqnarray}
g_{\mu\nu} &\rightarrow& \tilde g_{\mu\nu} = \Omega^2(x) g_{\mu\nu},  \quad
g^{\mu\nu} \rightarrow \tilde g^{\mu\nu} = \Omega^{-2}(x) g^{\mu\nu}, \quad
 \nonumber\\
\phi &\rightarrow& \tilde \phi = \Omega^{-1}(x) \phi, \quad
H \rightarrow \tilde H = \Omega^{-1}(x) H,  \quad
A_\mu \rightarrow \tilde A_\mu = A_\mu.
\label{Weyl transf}
\end{eqnarray}
%%%%%%%%%%%%%%%%%%%%%%%%%%%%%%%%%%%%%%%%%%%%%%%%%%%%%%%%%%%%%%%%%%%
Actually, it is straightforward to prove the Weyl invariance of ${\cal L}$ 
when we use the formulae $\sqrt{-g} = \Omega^{-4} \sqrt{- \tilde g}$ and
%**   Curvature %%%%%%%%%%%%%%%%%%%%%%%%%%%%%%%%%%%%%%%%%%%%%%%%%%%%%%%%%
\begin{eqnarray}
R = \Omega^2 ( \tilde R + 6 \tilde \Box f - 6 \tilde g^{\mu\nu} \partial_\mu f \partial_\nu f ),
\label{Curvature}
\end{eqnarray}
%%%%%%%%%%%%%%%%%%%%%%%%%%%%%%%%%%%%%%%%%%%%%%%%%%%%%%%%%%%%%%%%%%%
with being  defined as $f = \log \Omega$ and $\tilde \Box f = \frac{1}{\sqrt{- \tilde g}} 
\partial_\mu (\sqrt{- \tilde g} \tilde g^{\mu\nu} \partial_\nu f) = \tilde g^{\mu\nu} 
\tilde \nabla_\mu \tilde \nabla_\nu f$.

As a gauge condition for the Weyl symmetry, two gauge conditions, those are, the Einstein gauge (E-gauge) and
the Jordan gauge (J-gauge), are usually taken in \cite{Kallosh1}-\cite{Linde2}. First, we shall take the Einstein gauge and investigate 
its implication.

With the unitary gauge $H(x) = \frac{1}{\sqrt{2}} e^{i \alpha \theta(x)} (0, h(x))^T$ where $\alpha$, $\theta(x)$ and 
$h(x)$ are respectively a real number, the Nambu-Goldstone boson and the physical Higgs field, 
the Einstein gauge (E-gauge) is of the form
%**   E-gauge %%%%%%%%%%%%%%%%%%%%%%%%%%%%%%%%%%%%%%%%%%%%%%%%%%%%%%%%%
\begin{eqnarray}
\phi^2 - 2 |H|^2 = \phi^2 - h^2 = 6.      
\label{E-gauge}
\end{eqnarray}
%%%%%%%%%%%%%%%%%%%%%%%%%%%%%%%%%%%%%%%%%%%%%%%%%%%%%%%%%%%%%%%%%%%
This $SO(1,1)$ invariant gauge choice can be parametrized in terms of a canonically normalized
real scalar field $\varphi$ as
%**   Parametrization  %%%%%%%%%%%%%%%%%%%%%%%%%%%%%%%%%%%%%%%%%%%%%%%%%%%%%%%%%
\begin{eqnarray}
\phi = \sqrt{6} \cosh \frac{\varphi}{\sqrt{6}},   \quad
h = \sqrt{6} \sinh \frac{\varphi}{\sqrt{6}}.
\label{Parametrization}
\end{eqnarray}
%%%%%%%%%%%%%%%%%%%%%%%%%%%%%%%%%%%%%%%%%%%%%%%%%%%%%%%%%%%%%%%%%%%
With the E-gauge (and the unitary gauge), the Lagrangian (\ref{Lagr 1}) can be rewritten as
%**   Lagr-E  %%%%%%%%%%%%%%%%%%%%%%%%%%%%%%%%%%%%%%%%%%%%%%%%%%%%%%%%%
\begin{eqnarray}
\frac{1}{\sqrt{-g}}{\cal L} &=&  \frac{1}{2} R - \frac{1}{2} g^{\mu\nu} 
\partial_\mu \varphi \partial_\nu \varphi - 3 g^2 g^{\mu\nu} B_\mu B_\nu \sinh^2 \frac{\varphi}{\sqrt{6}}   \nonumber\\
&-& \frac{1}{4} g^{\mu\nu} g^{\rho\sigma} F_{\mu\rho} F_{\nu\sigma} - V(\varphi),
\label{Lagr-E}
\end{eqnarray}
%%%%%%%%%%%%%%%%%%%%%%%%%%%%%%%%%%%%%%%%%%%%%%%%%%%%%%%%%%%%%%%%%%%
where $V(\varphi)$ is the potential term $V(\phi, H)$ substituted by Eq. (\ref{Parametrization}).
And we have defined a new gauge field $B_\mu$ as $B_\mu = A_\mu + \frac{\alpha}{g} \partial_\mu \theta$
and consequently the gauge strength is now given by $F_{\mu\nu} \equiv \partial_\mu A_\nu - \partial_\nu A_\mu
= \partial_\mu B_\nu - \partial_\nu B_\mu$. Note that the third term in the RHS of Eq. (\ref{Lagr-E}) is written as
$\frac{1}{2} g^2 g^{\mu\nu} B_\mu B_\nu h^2$ which corresponds to the mass term of the gauge field in the
conventional formulation when the spontaneous symmetry breakdown (SSB) happens.  In other words,
the gauge field $A_\mu$ eats the Nambu-Goldstone boson $\theta$, thereby becoming massive after the SSB.
However, it is not $h$ but $\varphi$ that one now regards as a fundamental field,
so the interpretation of the third term as the Higgs mass term is not suitable within the present framework.

Next, let us take the Jordan gauge (J-gauge) instead of the E-gauge. With the J-gauge $\phi = \sqrt{6}$,
the Lagrangian (\ref{Lagr 1}) is reduced to
%**   Lagr-J  %%%%%%%%%%%%%%%%%%%%%%%%%%%%%%%%%%%%%%%%%%%%%%%%%%%%%%%%%
\begin{eqnarray}
\frac{1}{\sqrt{-g}}{\cal L} =  \frac{1}{2} \left( 1 - \frac{|H|^2}{3} \right) R  - g^{\mu\nu} (D_\mu H)^\dagger (D_\nu H)
- \frac{1}{4} g^{\mu\nu} g^{\rho\sigma} F_{\mu\rho} F_{\nu\sigma} - V(H),
\label{Lagr-J}
\end{eqnarray}
%%%%%%%%%%%%%%%%%%%%%%%%%%%%%%%%%%%%%%%%%%%%%%%%%%%%%%%%%%%%%%%%%%%
where $V(H) \equiv V(\phi = \sqrt{6}, H)$. In order to show the equivalence of the Lagrangian in between the E-gauge and
the J-gauge, one can make use of a scale transformation in Eq.  (\ref{Weyl transf}) by putting 
%**   Omega  %%%%%%%%%%%%%%%%%%%%%%%%%%%%%%%%%%%%%%%%%%%%%%%%%%%%%%%%%
\begin{eqnarray}
\Omega =  1 - \frac{|H|^2}{3} = 1 - \frac{h^2}{6},
\label{Omega}
\end{eqnarray}
%%%%%%%%%%%%%%%%%%%%%%%%%%%%%%%%%%%%%%%%%%%%%%%%%%%%%%%%%%%%%%%%%%%
where the unitary gauge is used in the second equality. After a straightforward calculation, we obtain the following
Lagrangian:
%**   Lagr-J2  %%%%%%%%%%%%%%%%%%%%%%%%%%%%%%%%%%%%%%%%%%%%%%%%%%%%%%%%%
\begin{eqnarray}
\frac{1}{\sqrt{- g}}{\cal L} &=&  \frac{1}{2} R  - \frac{1}{2 \Omega^4} g^{\mu\nu} \partial_\mu h \partial_\nu h
- \frac{1}{2 \Omega^2} g^2 g^{\mu\nu} B_\mu B_\nu h^2               \nonumber\\
&-& \frac{1}{4} g^{\mu\nu} g^{\rho\sigma} F_{\mu\rho} F_{\nu\sigma} - \Omega^{-4} V(H),
\label{Lagr-J2}
\end{eqnarray}
%%%%%%%%%%%%%%%%%%%%%%%%%%%%%%%%%%%%%%%%%%%%%%%%%%%%%%%%%%%%%%%%%%%
where we have omitted to write tildes on fields denoting the transformed fields for simplicity of the
presentation.
Moreover, introducing $h' = \frac{1}{h}$ and then defining the canonically normalized scalar field $\varphi$
in terms of a differential equation 
%**   Diff-eq  %%%%%%%%%%%%%%%%%%%%%%%%%%%%%%%%%%%%%%%%%%%%%%%%%%%%%%%%%
\begin{eqnarray}
\frac{d \varphi}{d h'} =  \frac{1}{h'^2 - \frac{1}{6}},
\label{Diff-eq}
\end{eqnarray}
%%%%%%%%%%%%%%%%%%%%%%%%%%%%%%%%%%%%%%%%%%%%%%%%%%%%%%%%%%%%%%%%%%%
which is simply solved to     
%**   Diff-eq-sol  %%%%%%%%%%%%%%%%%%%%%%%%%%%%%%%%%%%%%%%%%%%%%%%%%%%%%%%%%
\begin{eqnarray}
h' \equiv \frac{1}{h} =  \frac{1}{\sqrt{6} \tanh \frac{\varphi}{\sqrt{6}}},
\label{Diff-eq-sol}
\end{eqnarray}
%%%%%%%%%%%%%%%%%%%%%%%%%%%%%%%%%%%%%%%%%%%%%%%%%%%%%%%%%%%%%%%%%%%
the Lagrangian (\ref{Lagr-J2}) can be cast to 
%**   Lagr-J3  %%%%%%%%%%%%%%%%%%%%%%%%%%%%%%%%%%%%%%%%%%%%%%%%%%%%%%%%%
\begin{eqnarray}
\frac{1}{\sqrt{- g}}{\cal L} &=&  \frac{1}{2} R  - \frac{1}{2} g^{\mu\nu} \partial_\mu \varphi \partial_\nu \varphi
- 3 g^2 g^{\mu\nu} B_\mu B_\nu \sinh^2 \frac{\varphi}{\sqrt{6}}               \nonumber\\
&-& \frac{1}{4} g^{\mu\nu} g^{\rho\sigma} F_{\mu\rho} F_{\nu\sigma} - \Omega^{-4} V(H).
\label{Lagr-J3}
\end{eqnarray}
%%%%%%%%%%%%%%%%%%%%%%%%%%%%%%%%%%%%%%%%%%%%%%%%%%%%%%%%%%%%%%%%%%%

Finally, to complete the proof of equivalence, we must specify the form of the potential $V(\phi, H)$. As the most general quadratic expression
which preserves the Weyl invariance and reduces to the Higgs potential in the low energy, let us consider the following potential:
%**   Potential  %%%%%%%%%%%%%%%%%%%%%%%%%%%%%%%%%%%%%%%%%%%%%%%%%%%%%%%%%
\begin{eqnarray}
V(\phi, H) =  \frac{\lambda}{36} F(z) \left[ 2 |H|^2 - G(z) \phi^2 \right]^2,
\label{Potential}
\end{eqnarray}
%%%%%%%%%%%%%%%%%%%%%%%%%%%%%%%%%%%%%%%%%%%%%%%%%%%%%%%%%%%%%%%%%%%
where $z$ is a gauge-invariant quantity defined as 
%**   z  %%%%%%%%%%%%%%%%%%%%%%%%%%%%%%%%%%%%%%%%%%%%%%%%%%%%%%%%%
\begin{eqnarray}
z \equiv  \frac{\sqrt{2 |H|^2}}{\phi} = \frac{h}{\phi} = \tanh \frac{\varphi}{\sqrt{6}},
\label{z}
\end{eqnarray}
%%%%%%%%%%%%%%%%%%%%%%%%%%%%%%%%%%%%%%%%%%%%%%%%%%%%%%%%%%%%%%%%%%%
where at the last equality, the J-gauge and Eq. (\ref{Diff-eq-sol}) are used.

In the E-gauge, the potential term is written as
%**   Potential-E  %%%%%%%%%%%%%%%%%%%%%%%%%%%%%%%%%%%%%%%%%%%%%%%%%%%%%%%%%
\begin{eqnarray}
V(\varphi) =  \lambda F(z) \left[ \sinh^2 \frac{\varphi}{\sqrt{6}} 
- G(z) \cosh^2 \frac{\varphi}{\sqrt{6}} \right]^2,
\label{Potential-E}
\end{eqnarray}
%%%%%%%%%%%%%%%%%%%%%%%%%%%%%%%%%%%%%%%%%%%%%%%%%%%%%%%%%%%%%%%%%%%
where $z = \tanh \frac{\varphi}{\sqrt{6}}$. On the other hand, the potential
in the J-gauge takes the form 
%**   Potential-J  %%%%%%%%%%%%%%%%%%%%%%%%%%%%%%%%%%%%%%%%%%%%%%%%%%%%%%%%%
\begin{eqnarray}
\Omega^{-4} V(H) &=& \cosh^4 \frac{\varphi}{\sqrt{6}} \cdot \frac{\lambda}{36}
F(z) \left[ 6 \tanh^2 \frac{\varphi}{\sqrt{6}} 
- 6 G(z) \right]^2     \nonumber\\
&=& \lambda F(z) \left[ \sinh^2 \frac{\varphi}{\sqrt{6}} 
- G(z) \cosh^2 \frac{\varphi}{\sqrt{6}} \right]^2,
\label{Potential-J}
\end{eqnarray}
%%%%%%%%%%%%%%%%%%%%%%%%%%%%%%%%%%%%%%%%%%%%%%%%%%%%%%%%%%%%%%%%%%%
which is equivalent to the potential (\ref{Potential-E}) in the E-gauge,
so we have completed the proof of the equivalence of the Lagrangian in both the gauge conditions.

%%%%%%%%%%%%%%%%%%%%%%%%%%%%%%%%%%%%%%%%%%%%%%%%%%%%%%%%%%%%%%%%%%%%%
%%%%%%%%%%%%%%%%%%%%%%%%%%%%%%   SEC  3    %%%%%%%%%%%%%%%%%%%%%%%%%%
%%%%%%%%%%%%%%%%%%%%%%%%%%%%%%%%%%%%%%%%%%%%%%%%%%%%%%%%%%%%%%%%%%%%%
\section{SO(1,1) accidental symmetry and shift symmetry}

For explanation of a relation between an $SO(1,1)$ global symmetry and a shift symmetry,
let us consider the simplest $SO(1,1)$ invariant potential
%**   Sim-Potential  %%%%%%%%%%%%%%%%%%%%%%%%%%%%%%%%%%%%%%%%%%%%%%%%%%%%%%%%%
\begin{eqnarray}
V_0 (\phi, H) =  \frac{\lambda}{36} \left( 2 |H|^2 - \phi^2 \right)^2
= \frac{\lambda}{36} \left( h^2 - \phi^2 \right)^2.
\label{Sim-Potential}
\end{eqnarray}
%%%%%%%%%%%%%%%%%%%%%%%%%%%%%%%%%%%%%%%%%%%%%%%%%%%%%%%%%%%%%%%%%%%
Moreover, we switch off the gauge field, $A_\mu = 0$. Then, in the unitary gauge,
the Lagrangian (\ref{Lagr 1}) is of form 
%**   Lagr-SO  %%%%%%%%%%%%%%%%%%%%%%%%%%%%%%%%%%%%%%%%%%%%%%%%%%%%%%%%%
\begin{eqnarray}
\frac{1}{\sqrt{-g}}{\cal L} =  \frac{1}{12} \left( \phi^2 - h^2 \right) R  
+ \frac{1}{2} g^{\mu\nu} \partial_\mu \phi \partial_\nu \phi
- \frac{1}{2} g^{\mu\nu} \partial_\mu h \partial_\nu h - V_0 (\phi, H).
\label{Lagr-SO}
\end{eqnarray}
%%%%%%%%%%%%%%%%%%%%%%%%%%%%%%%%%%%%%%%%%%%%%%%%%%%%%%%%%%%%%%%%%%%
It is obvious that there is a global $SO(1,1)$ symmetry in this Lagrangian. 
Namely, the Lagrangian (\ref{Lagr-SO}) is invariant under the global $SO(1,1)$
transformation ($\varphi_0$ is a constant parameter)
%**   SO(1,1)  %%%%%%%%%%%%%%%%%%%%%%%%%%%%%%%%%%%%%%%%%%%%%%%%%%%%%%%%%
\begin{eqnarray}
\left(
    \begin{array}{c}
      \phi' \\
      h' \\
    \end{array}
  \right)
= \left(
    \begin{array}{cc}
      \cosh \frac{\varphi_0}{\sqrt{6}} & \sinh \frac{\varphi_0}{\sqrt{6}}   \\
      \sinh \frac{\varphi_0}{\sqrt{6}} & \cosh \frac{\varphi_0}{\sqrt{6}}  \\
    \end{array}
  \right)
  \left(
    \begin{array}{c}
      \phi \\
      h \\
    \end{array}
  \right).
\label{SO(1,1)}
\end{eqnarray}
%%%%%%%%%%%%%%%%%%%%%%%%%%%%%%%%%%%%%%%%%%%%%%%%%%%%%%%%%%%%%%%%%%%
Using the parametrization (\ref{Parametrization}) in the E-gauge which respects the $SO(1,1)$ symmetry, 
this $SO(1,1)$ transformation can be written as
%**   SO(1,1)-E  %%%%%%%%%%%%%%%%%%%%%%%%%%%%%%%%%%%%%%%%%%%%%%%%%%%%%%%%%
\begin{eqnarray}
\left(
    \begin{array}{c}
      \phi' \\
      h' \\
    \end{array}
  \right)
\equiv
\left(
    \begin{array}{c}
      \sqrt{6}  \cosh \frac{\varphi'}{\sqrt{6}} \\
      \sqrt{6}  \sinh \frac{\varphi'}{\sqrt{6}} \\
    \end{array}
  \right)
= \left(
    \begin{array}{c}
      \sqrt{6}  \cosh \frac{\varphi + \varphi_0}{\sqrt{6}} \\
      \sqrt{6}  \sinh \frac{\varphi + \varphi_0}{\sqrt{6}} \\
    \end{array}
  \right), 
\label{SO(1,1)-E}
\end{eqnarray}
%%%%%%%%%%%%%%%%%%%%%%%%%%%%%%%%%%%%%%%%%%%%%%%%%%%%%%%%%%%%%%%%%%%
where we have used addition theorem of hyperbolic functions
%**   Addition  %%%%%%%%%%%%%%%%%%%%%%%%%%%%%%%%%%%%%%%%%%%%%%%%%%%%%%%%%
\begin{eqnarray}
\sinh (\alpha \pm \beta) &=& \sinh \alpha \cosh \beta \pm \cosh \alpha \sinh \beta,    \nonumber\\
\cosh (\alpha \pm \beta) &=& \cosh \alpha \cosh \beta \pm \sinh \alpha \sinh \beta.
\label{Addition}
\end{eqnarray}
%%%%%%%%%%%%%%%%%%%%%%%%%%%%%%%%%%%%%%%%%%%%%%%%%%%%%%%%%%%%%%%%%%%
Eq. (\ref{SO(1,1)-E}) clearly means that the $SO(1,1)$ transformation is nothing but a shift transformation
of $\varphi$ in the E-gauge, that is,
%**   Shift  %%%%%%%%%%%%%%%%%%%%%%%%%%%%%%%%%%%%%%%%%%%%%%%%%%%%%%%%%
\begin{eqnarray}
\varphi \rightarrow \varphi' = \varphi + \varphi_0.
\label{Shift}
\end{eqnarray}
%%%%%%%%%%%%%%%%%%%%%%%%%%%%%%%%%%%%%%%%%%%%%%%%%%%%%%%%%%%%%%%%%%%

In inflation models, up to derivative terms, the potential term which is invariant under the shift symmetry
is known to be only a cosmological constant. Actually, the Lagrangian (\ref{Lagr-SO}) is written in the
E-gauge as  
%**   Lagr-SO-E  %%%%%%%%%%%%%%%%%%%%%%%%%%%%%%%%%%%%%%%%%%%%%%%%%%%%%%%%%
\begin{eqnarray}
\frac{1}{\sqrt{-g}}{\cal L} =  \frac{1}{2} R  
- \frac{1}{2} g^{\mu\nu} \partial_\mu \varphi \partial_\nu \varphi
- \lambda.
\label{Lagr-SO-E}
\end{eqnarray}
%%%%%%%%%%%%%%%%%%%%%%%%%%%%%%%%%%%%%%%%%%%%%%%%%%%%%%%%%%%%%%%%%%%
From this observation, it turns out that the existence of the non-trivial potential term for inflation
demands that one needs to have the potential which is not invariant under the $SO(1,1)$ transformation.

However, when we incorporate the gauge field in the Lagrangian as in the present formalism, the situation
completely changes, that is, the gauge field couples to only the Higgs field, thereby breaking the $SO(1,1)$
symmetry explicitly. In this sense, the $SO(1,1)$ symmetry is an "accidental" symmetry which exists only
in the gravitational sector. The accidental symmetry or custodial symmetry has some important consequences
in the standard model and a theory beyond standard model \cite{Logan}. In the present theory, however, it is 
not clear how this accidental symmetry leads to important consequences.

%%%%%%%%%%%%%%%%%%%%%%%%%%%%%%%%%%%%%%%%%%%%%%%%%%%%%%%%%%%%%%%%%%%%%
%%%%%%%%%%%%%%%%%%%%%%%%%%%%%%   SEC  4    %%%%%%%%%%%%%%%%%%%%%%%%%%
%%%%%%%%%%%%%%%%%%%%%%%%%%%%%%%%%%%%%%%%%%%%%%%%%%%%%%%%%%%%%%%%%%%%%
\section{Higgs mechanism}

In this section, we wish to take account of the Higgs mechanism in the formalism at hand. For generality,
let us consider the most general form of the potential, Eq. (\ref{Potential}), in the Einstein gauge.

If we impose boundary conditions on the functions $F(z)$ and $G(z)$ in a small field limit
%**   BC1  %%%%%%%%%%%%%%%%%%%%%%%%%%%%%%%%%%%%%%%%%%%%%%%%%%%%%%%%%
\begin{eqnarray}
\lim_{z \to 0} F(z) = 9,  \quad \lim_{z \to 0} G(z) = \frac{v^2}{6},
\label{BC1}
\end{eqnarray}
%%%%%%%%%%%%%%%%%%%%%%%%%%%%%%%%%%%%%%%%%%%%%%%%%%%%%%%%%%%%%%%%%%%
the potential (\ref{Potential}) takes the following form in the limit $z \rightarrow 0$, 
or equivalently $\varphi \rightarrow 0$:
%**   Potential-S  %%%%%%%%%%%%%%%%%%%%%%%%%%%%%%%%%%%%%%%%%%%%%%%%%%%%%%%%%
\begin{eqnarray}
V \simeq \frac{\lambda}{4} ( \varphi^2 - v^2 )^2 \simeq \frac{\lambda}{4} ( h^2 - v^2 )^2,
\label{Potential-S}
\end{eqnarray}
%%%%%%%%%%%%%%%%%%%%%%%%%%%%%%%%%%%%%%%%%%%%%%%%%%%%%%%%%%%%%%%%%%%
where we have used $\varphi \simeq h$ for $\varphi \ll 1$. This is the conventional Higgs potential, so when
the Higgs field takes the vacuum expectation value $< h > = v$, the gauge field $B_\mu$
has the mass $M_B = g v$ as seen in Eq. (\ref{Lagr-E}). Note that $\phi \simeq \sqrt{6}$ for $\varphi \ll 1$,
which implies that the dynamical degree of freedom associated with $\phi$ naturally disappears and is fixed to
be a definite value of the J-gauge in a small field limit.

In a large field limit $\varphi \gg 1$, the term $3 g^2 B_\mu^2 \sinh^2 \frac{\varphi}{\sqrt{6}}$, giving 
rise to the mass of the gauge field in a small field limit, loses a role as the mass term, and becomes an
interaction term between the gauge field $B_\mu$ and the inflaton $\varphi$.
In fact, since for $\varphi \gg 1$, Eq. (\ref{Parametrization}) gives us
%**   Phi-h  %%%%%%%%%%%%%%%%%%%%%%%%%%%%%%%%%%%%%%%%%%%%%%%%%%%%%%%%%
\begin{eqnarray}
\phi \simeq h \simeq \sqrt{\frac{3}{2}} \ e^{\frac{\varphi}{\sqrt{6}}},
\label{Phi-h}
\end{eqnarray}
%%%%%%%%%%%%%%%%%%%%%%%%%%%%%%%%%%%%%%%%%%%%%%%%%%%%%%%%%%%%%%%%%%%
we obtain
%**   Non-renorma  %%%%%%%%%%%%%%%%%%%%%%%%%%%%%%%%%%%%%%%%%%%%%%%%%%%%%%%%%
\begin{eqnarray}
3 g^2 B_\mu^2 \sinh^2 \frac{\varphi}{\sqrt{6}} = \frac{1}{2} g^2 B_\mu^2 h^2 
\simeq \frac{3}{4} g^2 B_\mu^2 e^{\sqrt{\frac{2}{3}} \varphi}.
\label{Non-renorma}
\end{eqnarray}
%%%%%%%%%%%%%%%%%%%%%%%%%%%%%%%%%%%%%%%%%%%%%%%%%%%%%%%%%%%%%%%%%%%
This term gives rise to non-renormalizable interactions between the gauge field and the inflaton,
and plays a role in the process of reheating \cite{Tomoyose}.

%%%%%%%%%%%%%%%%%%%%%%%%%%%%%%%%%%%%%%%%%%%%%%%%%%%%%%%%%%%%%%%%%%%%%
%%%%%%%%%%%%%%%%%%%%%%%%%%%%%%   SEC  5    %%%%%%%%%%%%%%%%%%%%%%%%%%
%%%%%%%%%%%%%%%%%%%%%%%%%%%%%%%%%%%%%%%%%%%%%%%%%%%%%%%%%%%%%%%%%%%%%
\section{The generalization to non-abelian gauge field}

For completeness, let us extend the abelian gauge field to the non-abelian gauge field
since the standard model is constructed on the basis of the gauge group $SU(3)_C \times SU(1)_L \times U(1)_Y$. For clarity, 
we shall consider only the $SU(2)$ gauge group  since the generalization to a general non-abelian 
gauge field is straightforward.

Let us start with the $SU(2)$ generalization of the Lagrangian (\ref{Lagr 1})
%**   NA-Lagr 1  %%%%%%%%%%%%%%%%%%%%%%%%%%%%%%%%%%%%%%%%%%%%%%%%%%%%%%%%%
\begin{eqnarray}
\frac{1}{\sqrt{-g}}{\cal L} &=&  \frac{1}{12} \left( \phi^2 - 2 |H|^2 \right) R + \frac{1}{2} g^{\mu\nu} 
\partial_\mu \phi \partial_\nu \phi - g^{\mu\nu} (D_\mu H)^\dagger (D_\nu H)    \nonumber\\
&-& \frac{1}{4} g^{\mu\nu} g^{\rho\sigma} F_{\mu\rho}^a F_{\nu\sigma}^a - V(\phi, H),
\label{NA-Lagr 1}
\end{eqnarray}
%%%%%%%%%%%%%%%%%%%%%%%%%%%%%%%%%%%%%%%%%%%%%%%%%%%%%%%%%%%%%%%%%%%
where $a$ is an $SU(2)$ index running over $1, 2, 3$, and the covariant derivative and field strength 
are respectively defined as 
%**   NA-Def 1  %%%%%%%%%%%%%%%%%%%%%%%%%%%%%%%%%%%%%%%%%%%%%%%%%%%%%%%%%
\begin{eqnarray}
D_\mu H &=& (\partial_\mu - i g \tau^a A_\mu^a) H,   \quad
(D_\mu H)^\dagger =H^\dagger (\overleftarrow{\partial}_\mu + i g \tau^a A_\mu^a),   \quad \nonumber\\
F_{\mu\nu}^a &=& \partial_\mu A_\nu^a - \partial_\nu A_\mu^a + g \varepsilon^{abc} A_\mu^b A_\nu^c.
\label{NA-Def 1}
\end{eqnarray}
%%%%%%%%%%%%%%%%%%%%%%%%%%%%%%%%%%%%%%%%%%%%%%%%%%%%%%%%%%%%%%%%%%%
Here $g$ is an $SU(2)$ coupling constant. Furthermore, the matrices $\tau^a$ are defined as half of the Pauli ones, i.e.,
$\tau^a = \frac{1}{2} \sigma^a$, so the following relations are satisfied:
%**   tau-matrix  %%%%%%%%%%%%%%%%%%%%%%%%%%%%%%%%%%%%%%%%%%%%%%%%%%%%%%%%%
\begin{eqnarray}
\{ \tau^a, \tau^b \} = \frac{1}{2} \delta^{ab},  \quad
[ \tau^a, \tau^b ] = i \varepsilon^{abc} \tau^c.
\label{tau-matrix}
\end{eqnarray}
%%%%%%%%%%%%%%%%%%%%%%%%%%%%%%%%%%%%%%%%%%%%%%%%%%%%%%%%%%%%%%%%%%%

In order to see the Higgs mechanism discussed in the previous section explicitly, it is
convenient to go to the unitary gauge. To do that, we first parametrize the Higgs doublet as
%**   U-gauge  %%%%%%%%%%%%%%%%%%%%%%%%%%%%%%%%%%%%%%%%%%%%%%%%%%%%%%%%%
\begin{eqnarray}
H(x) = \frac{1}{\sqrt{2}} U^{-1}(x) \left(
    \begin{array}{c}
      0 \\
      h(x) \\
    \end{array}
  \right),
\label{U-gauge}
\end{eqnarray}
%%%%%%%%%%%%%%%%%%%%%%%%%%%%%%%%%%%%%%%%%%%%%%%%%%%%%%%%%%%%%%%%%%%
where a unitary matrix $U(x)$ is defined as $U(x) = e^{ -i \alpha \tau^a \theta^a(x)}$ with $\alpha$ and
$\theta^a(x)$ being a real number and the Nambu-Goldstone fields, respectively.
Then, we will define new fields in the unitary gauge by
%**   NA-new fields  %%%%%%%%%%%%%%%%%%%%%%%%%%%%%%%%%%%%%%%%%%%%%%%%%%%%%%%%%
\begin{eqnarray}
H^u(x) &=& U(x) H(x) = \frac{1}{\sqrt{2}} \left(
    \begin{array}{c}
      0 \\
      h(x) \\
    \end{array}
  \right),    \nonumber\\
\tau^a B_\mu^a &=& U(x) \tau^a A_\mu^a U^{-1}(x) - \frac{i}{g} \partial_\mu U(x) U^{-1}(x).
\label{NA-new fields}
\end{eqnarray}
%%%%%%%%%%%%%%%%%%%%%%%%%%%%%%%%%%%%%%%%%%%%%%%%%%%%%%%%%%%%%%%%%%%
Using these new fields, after an easy calculation, we find the following relations
%**   NA-Rel 1  %%%%%%%%%%%%%%%%%%%%%%%%%%%%%%%%%%%%%%%%%%%%%%%%%%%%%%%%%
\begin{eqnarray}
D_\mu H = U^{-1}(x) D_\mu H^u,   \quad
F_{\mu\nu}^a F^{a \mu\nu} = F_{\mu\nu}^a(B) F^{a \mu\nu}(B),
\label{NA-Rel 1}
\end{eqnarray}
%%%%%%%%%%%%%%%%%%%%%%%%%%%%%%%%%%%%%%%%%%%%%%%%%%%%%%%%%%%%%%%%%%%
where $D_\mu H^u$ and $F_{\mu\nu}^a(B)$ are respectively defined as
%**   NA-Def 2  %%%%%%%%%%%%%%%%%%%%%%%%%%%%%%%%%%%%%%%%%%%%%%%%%%%%%%%%%
\begin{eqnarray}
D_\mu H^u = (\partial_\mu - i g \tau^a B_\mu^a) H^u,   \quad
F_{\mu\nu}^a(B) = \partial_\mu B_\nu^a - \partial_\nu B_\mu^a + g \varepsilon^{abc} B_\mu^b B_\nu^c.
\label{NA-Def 2}
\end{eqnarray}
%%%%%%%%%%%%%%%%%%%%%%%%%%%%%%%%%%%%%%%%%%%%%%%%%%%%%%%%%%%%%%%%%%%

Thus, with the unitary gauge, the Lagrangian (\ref{NA-Lagr 1}) becomes
%**   NA-Lagr 1-U  %%%%%%%%%%%%%%%%%%%%%%%%%%%%%%%%%%%%%%%%%%%%%%%%%%%%%%%%%
\begin{eqnarray}
\frac{1}{\sqrt{-g}}{\cal L} &=&  \frac{1}{12} \left( \phi^2 - 2 |H^u|^2 \right) R + \frac{1}{2} g^{\mu\nu} 
\partial_\mu \phi \partial_\nu \phi - g^{\mu\nu} (D_\mu H^u)^\dagger (D_\nu H^u)    \nonumber\\
&-& \frac{1}{4} g^{\mu\nu} g^{\rho\sigma} F_{\mu\rho}^a(B) F_{\nu\sigma}^a(B) - V(\phi, U^{-1} H^u).
\label{NA-Lagr 1-U}
\end{eqnarray}
%%%%%%%%%%%%%%%%%%%%%%%%%%%%%%%%%%%%%%%%%%%%%%%%%%%%%%%%%%%%%%%%%%%
This Lagrangian is easily evaluated in the Einstein gauge to 
%**   NA-Lagr 1-E  %%%%%%%%%%%%%%%%%%%%%%%%%%%%%%%%%%%%%%%%%%%%%%%%%%%%%%%%%
\begin{eqnarray}
\frac{1}{\sqrt{-g}}{\cal L} &=&  \frac{1}{2} R - \frac{1}{2} g^{\mu\nu} 
\partial_\mu \varphi \partial_\nu \varphi - \frac{3 g^2}{4} g^{\mu\nu} B_\mu^a B_\nu^a \sinh^2 \frac{\varphi}{\sqrt{6}}
\nonumber\\
&-& \frac{1}{4} g^{\mu\nu} g^{\rho\sigma} F_{\mu\rho}^a(B) F_{\nu\sigma}^a(B) - V(\phi, U^{-1} H^u).
\label{NA-Lagr 1-E}
\end{eqnarray}
%%%%%%%%%%%%%%%%%%%%%%%%%%%%%%%%%%%%%%%%%%%%%%%%%%%%%%%%%%%%%%%%%%%
Here the potential is taken to be the most general quadratic form (\ref{Potential}) and is rewritten in the E-gauge as
%**   NA-Potential  %%%%%%%%%%%%%%%%%%%%%%%%%%%%%%%%%%%%%%%%%%%%%%%%%%%%%%%%%
\begin{eqnarray}
V(\phi, H) =  V(\phi, U^{-1} H^u) = V(\phi, H^u) 
= \lambda F(z) \left[ \sinh^2 \frac{\varphi}{\sqrt{6}} - G(z) \cosh^2 \frac{\varphi}{\sqrt{6}} \right]^2,
\label{NA-Potential}
\end{eqnarray}
%%%%%%%%%%%%%%%%%%%%%%%%%%%%%%%%%%%%%%%%%%%%%%%%%%%%%%%%%%%%%%%%%%% 
where $z \equiv \tanh \frac{\varphi}{\sqrt{6}}$. Then, with the boundary conditions (\ref{BC1}) this potential takes
the same form as (\ref{Potential-S}) in small field values so that when the Higgs field takes the vacuum expectation value 
$< h > = v$, the non-abelain gauge field $B_\mu$ has the mass $M_B = \frac{1}{2} g v$. Note that in the non-abelain
gauge field, the Higgs mass squared term depends on the inflaton field $\varphi$ like 
$\frac{3 g^2}{4} g^{\mu\nu} B_\mu^a B_\nu^a \sinh^2 \frac{\varphi}{\sqrt{6}}$, so its physical property has a perfectly
similar aspect to that in the abelian gauge field.

%%%%%%%%%%%%%%%%%%%%%%%%%%%%%%%%%%%%%%%%%%%%%%%%%%%%%%%%%%%%%%%%%%%%%
%%%%%%%%%%%%%%%%%%%%%%%%%%%%%%   SEC  6    %%%%%%%%%%%%%%%%%%%%%%%%%%
%%%%%%%%%%%%%%%%%%%%%%%%%%%%%%%%%%%%%%%%%%%%%%%%%%%%%%%%%%%%%%%%%%%%%
\section{Inflation}

Based on our abelian gauge model, we will turn our attention to inflation. In particular,
we wish to ask ourselves if the abelian model accounts for the recent BICEP2 results
or not. It will turn out that the answer is not affirmative even if we consider a rather singular potential
in addition to a non-singular one since our model shares a common feature with the Starobinsky
inflation model \cite{Starobinsky}.

As a first step, let us select a boundary condition in a large field limit
%**   BC2  %%%%%%%%%%%%%%%%%%%%%%%%%%%%%%%%%%%%%%%%%%%%%%%%%%%%%%%%%
\begin{eqnarray}
\lim_{z \to 1} F(z) = F(1) < \infty,  \quad \lim_{z \to 1} G(z) = 1,
\label{BC2}
\end{eqnarray}
%%%%%%%%%%%%%%%%%%%%%%%%%%%%%%%%%%%%%%%%%%%%%%%%%%%%%%%%%%%%%%%%%%%
where $F(1)$ is a finite quantity. The second boundary condition $G(1) = 1$ is chosen in a such way that 
our model does not lead to cyclic cosmology \cite{Bars1, Bars2}.   

In the Einstein gauge, the most general quadratic potential (\ref{Potential}) is written in
a concise form as
%**   Potential-Inf  %%%%%%%%%%%%%%%%%%%%%%%%%%%%%%%%%%%%%%%%%%%%%%%%%%%%%%%%%
\begin{eqnarray}
V(\phi, H) &=&  \frac{\lambda}{36} F(z) \left[ 2 |H|^2 - G(z) \phi^2 \right]^2  \nonumber\\
&=& \lambda F(z) \left[ \frac{z^2 - G(z)}{z^2 - 1} \right]^2,
\label{Potential-Inf}
\end{eqnarray}
%%%%%%%%%%%%%%%%%%%%%%%%%%%%%%%%%%%%%%%%%%%%%%%%%%%%%%%%%%%%%%%%%%%
where $z \equiv \frac{\sqrt{2 |H|^2}}{\phi} = \frac{h}{\phi} = \tanh \frac{\varphi}{\sqrt{6}}$.
However, this concise expression turns out to be not convenient for analysing cosmological parameters
at large field values, so instead we will use the following equivalent form of the potential
%**   Potential-Inf2  %%%%%%%%%%%%%%%%%%%%%%%%%%%%%%%%%%%%%%%%%%%%%%%%%%%%%%%%%
\begin{eqnarray}
V(\varphi) = \lambda F(\tanh \frac{\varphi}{\sqrt{6}}) \cdot \sinh^4 \frac{\varphi}{\sqrt{6}}
\left[ 1- G(\tanh \frac{\varphi}{\sqrt{6}}) \tanh^{-2} \frac{\varphi}{\sqrt{6}} \right]^2.
\label{Potential-Inf2}
\end{eqnarray}
%%%%%%%%%%%%%%%%%%%%%%%%%%%%%%%%%%%%%%%%%%%%%%%%%%%%%%%%%%%%%%%%%%%   
At large field values $\varphi \gg 1$, hyperbolic functions are approximated to
%**   Hyperbolic  %%%%%%%%%%%%%%%%%%%%%%%%%%%%%%%%%%%%%%%%%%%%%%%%%%%%%%%%%
\begin{eqnarray}
\sinh \frac{\varphi}{\sqrt{6}} \simeq \frac{1}{2} e^{\frac{\varphi}{\sqrt{6}}},  \quad
\tanh \frac{\varphi}{\sqrt{6}} \simeq 1 - 2 e^{- \sqrt{\frac{2}{3}} \varphi} 
\left( 1 - e^{- \sqrt{\frac{2}{3}} \varphi} \right).
\label{Hyperbolic}
\end{eqnarray}
%%%%%%%%%%%%%%%%%%%%%%%%%%%%%%%%%%%%%%%%%%%%%%%%%%%%%%%%%%%%%%%%%%%  

Next let us assume that $F(x)$ and $G(x)$ are analytic around $x = 1$. Then, for $\varphi \gg 1$
we obtain
%**   FG  %%%%%%%%%%%%%%%%%%%%%%%%%%%%%%%%%%%%%%%%%%%%%%%%%%%%%%%%%
\begin{eqnarray}
F(\tanh \frac{\varphi}{\sqrt{6}}) &\simeq& F(1) - 2 F'(1) e^{- \sqrt{\frac{2}{3}} \varphi} 
\left( 1 - e^{- \sqrt{\frac{2}{3}} \varphi} \right),   \nonumber\\
G(\tanh \frac{\varphi}{\sqrt{6}}) &\simeq& 1 - 2 G'(1) e^{- \sqrt{\frac{2}{3}} \varphi} 
\left( 1 - e^{- \sqrt{\frac{2}{3}} \varphi} \right),
\label{FG}
\end{eqnarray}
%%%%%%%%%%%%%%%%%%%%%%%%%%%%%%%%%%%%%%%%%%%%%%%%%%%%%%%%%%%%%%%%%%%  
where the boundary condition $G(1) = 1$ is used and $F'(x) \equiv \frac{d F(x)}{d x}$.
Using these approximation formulae holding at large field values, the potential (\ref{Potential-Inf2}) 
takes the form
%**   Potential-Inf3  %%%%%%%%%%%%%%%%%%%%%%%%%%%%%%%%%%%%%%%%%%%%%%%%%%%%%%%%%
\begin{eqnarray}
V(\varphi) \simeq A \left( 1- B e^{- \sqrt{\frac{2}{3}} \varphi} \right),
\label{Potential-Inf3}
\end{eqnarray}
%%%%%%%%%%%%%%%%%%%%%%%%%%%%%%%%%%%%%%%%%%%%%%%%%%%%%%%%%%%%%%%%%%%   
where we have defined
%**   AB  %%%%%%%%%%%%%%%%%%%%%%%%%%%%%%%%%%%%%%%%%%%%%%%%%%%%%%%%%
\begin{eqnarray}
A &=& \frac{\lambda}{4} F(1) \left( G'(1) - 2 \right)^2,   \nonumber\\
B &=& \frac{F'(1)}{F(1)} - \frac{3 G'(1) + 2}{G'(1) - 2}.
\label{AB}
\end{eqnarray}
%%%%%%%%%%%%%%%%%%%%%%%%%%%%%%%%%%%%%%%%%%%%%%%%%%%%%%%%%%%%%%%%%%%  
Here we assume that both $A$ and $B$ are positive and $G'(1) \neq 2$.

The potential (\ref{Potential-Inf3}) yields the slow-roll parameters
%**   SL parameters  %%%%%%%%%%%%%%%%%%%%%%%%%%%%%%%%%%%%%%%%%%%%%%%%%%%%%%%%%
\begin{eqnarray}
\epsilon_V &\equiv& \frac{1}{2} \left( \frac{V_{, \varphi}}{V} \right)^2
\simeq \frac{B^2}{3} e^{- 2 \sqrt{\frac{2}{3}} \varphi},   \nonumber\\
\eta_V &\equiv& \frac{V_{, \varphi\varphi}}{V} 
\simeq - \frac{2 B}{3} e^{- \sqrt{\frac{2}{3}} \varphi},
\label{SL parameters}
\end{eqnarray}
%%%%%%%%%%%%%%%%%%%%%%%%%%%%%%%%%%%%%%%%%%%%%%%%%%%%%%%%%%%%%%%%%%%  
where $V_{, \varphi} \equiv \frac{d V}{d \varphi}$ and $V_{, \varphi\varphi} 
\equiv \frac{d^2 V}{d \varphi^2}$. The number of e-folding can be evaluated to
%**   N  %%%%%%%%%%%%%%%%%%%%%%%%%%%%%%%%%%%%%%%%%%%%%%%%%%%%%%%%%
\begin{eqnarray}
N \equiv \int_{\varphi_e}^\varphi d \varphi \frac{1}{\sqrt{2 \epsilon_V}}
\simeq  \frac{3}{2B} \left( e^{\sqrt{\frac{2}{3}} \varphi} - e^{\sqrt{\frac{2}{3}} 
\varphi_e} \right).
\label{N}
\end{eqnarray}
%%%%%%%%%%%%%%%%%%%%%%%%%%%%%%%%%%%%%%%%%%%%%%%%%%%%%%%%%%%%%%%%%%% 
The inflation ends when the slow-roll parameters are approximately the unity, 
$\epsilon_V(\varphi_e) = 1$, so we obtain $e^{\sqrt{\frac{2}{3}} 
\varphi_e} = \frac{B}{\sqrt{3}}$. Then, we can derive the relation
holding $N \gg 1$
%**   e-varphi  %%%%%%%%%%%%%%%%%%%%%%%%%%%%%%%%%%%%%%%%%%%%%%%%%%%%%%%%%
\begin{eqnarray}
e^{\sqrt{\frac{2}{3}} \varphi} = \frac{2 B N}{3} + \frac{B}{\sqrt{3}}
\simeq \frac{2 B N}{3}.
\label{e-varphi}
\end{eqnarray}
%%%%%%%%%%%%%%%%%%%%%%%%%%%%%%%%%%%%%%%%%%%%%%%%%%%%%%%%%%%%%%%%%%% 
Using this relation (\ref{e-varphi}), the slow-roll parameters can be
expressed in terms of the number of e-folding like
%**   SL parameters 2 %%%%%%%%%%%%%%%%%%%%%%%%%%%%%%%%%%%%%%%%%%%%%%%%%%%%%%%%%
\begin{eqnarray}
\epsilon_V = \frac{3}{4 N^2}, \quad
\eta_V = - \frac{1}{N}.
\label{SL parameters 2}
\end{eqnarray}
%%%%%%%%%%%%%%%%%%%%%%%%%%%%%%%%%%%%%%%%%%%%%%%%%%%%%%%%%%%%%%%%%%%  
Then, the tensor-to-scalar ratio $r$ and the spectral index $n_s$ are
given by 
%**   Index %%%%%%%%%%%%%%%%%%%%%%%%%%%%%%%%%%%%%%%%%%%%%%%%%%%%%%%%%
\begin{eqnarray}
r \equiv 16 \epsilon_V = \frac{12}{N^2}, \quad
n_s \equiv 1 - 6 \epsilon_V + 2 \eta_V \simeq 1 - \frac{2}{N},
\label{Index}
\end{eqnarray}
%%%%%%%%%%%%%%%%%%%%%%%%%%%%%%%%%%%%%%%%%%%%%%%%%%%%%%%%%%%%%%%%%%% 
which are in a perfect agreement with the values obtained in the Starobinsky model \cite{Starobinsky}.  
Concretely, for $N = 60$, they take the values $r \simeq 0.003$ and $n_s \simeq 0.967$,
which do not coincide with the BICEP2 results $r \simeq 0.16$ and $n_s \simeq 0.96$. 

So far, we have considered only analytic functions for $F$ and $G$. One possible
extension is to consider non-analytic functions and check if they lead to 
the BICEP2 results or not.  A general treatment of non-analytic functions is
not easy, but an instance turns out to be enough to derive a general feature of the 
non-analytic functions.

Let us focus on an example of non-analytic functions such that
%**   Non-FG %%%%%%%%%%%%%%%%%%%%%%%%%%%%%%%%%%%%%%%%%%%%%%%%%%%%%%%%%
\begin{eqnarray}
F(z) = 10 - (1 - z^n)^\alpha, \quad
G(z) = \frac{v^2}{6} + \left(1 - \frac{v^2}{6} \right) z^m,
\label{Non-FG}
\end{eqnarray}
%%%%%%%%%%%%%%%%%%%%%%%%%%%%%%%%%%%%%%%%%%%%%%%%%%%%%%%%%%%%%%%%%%%  
where $0 < \alpha < 1$. Note that these specific functions satisfy the boundary conditions
(\ref{BC1}) and (\ref{BC2}), but the derivatives of $F$ at $z=1$ diverge, $F'(1) = \infty, 
F''(1) = \infty, \cdots $. Using a behavior of $F$ for $\varphi \gg 1$
%**   Non-F %%%%%%%%%%%%%%%%%%%%%%%%%%%%%%%%%%%%%%%%%%%%%%%%%%%%%%%%%
\begin{eqnarray}
F(\varphi) \simeq 10 - (2n)^\alpha e^{- \alpha \sqrt{\frac{2}{3}} \varphi},
\label{Non-F}
\end{eqnarray}
%%%%%%%%%%%%%%%%%%%%%%%%%%%%%%%%%%%%%%%%%%%%%%%%%%%%%%%%%%%%%%%%%%%  
the potential reads in a large field limit $\varphi \gg 1$
%**   Non-V %%%%%%%%%%%%%%%%%%%%%%%%%%%%%%%%%%%%%%%%%%%%%%%%%%%%%%%%%
\begin{eqnarray}
V(\varphi) \simeq \frac{5 \lambda}{2} (m-2)^2 \left[ 1 - \frac{(2n)^\alpha}{10} 
e^{- \alpha \sqrt{\frac{2}{3}} \varphi} \right],
\label{Non-V}
\end{eqnarray}
%%%%%%%%%%%%%%%%%%%%%%%%%%%%%%%%%%%%%%%%%%%%%%%%%%%%%%%%%%%%%%%%%%%  
where a term involving the factor $e^{- \alpha \sqrt{\frac{2}{3}} \varphi}$ in the non-analytic
function $F(\varphi)$ becomes more dominant compared to terms including the factor
$e^{- \sqrt{\frac{2}{3}} \varphi}$ coming from hyperbolic functions $\sinh \frac{\varphi}{\sqrt{6}}$
and $\tanh \frac{\varphi}{\sqrt{6}}$.  

Along the same line of argument as in the potential (\ref{Potential-Inf3}), it is straightforward
to derive the slow-roll parameters and cosmological indices whose result is summarized as
%**   Indices %%%%%%%%%%%%%%%%%%%%%%%%%%%%%%%%%%%%%%%%%%%%%%%%%%%%%%%%%
\begin{eqnarray}
\epsilon_V &=& \frac{3}{4 \alpha^2 N^2}, \quad
\eta_V = - \frac{1}{N}, \nonumber\\
r &=& \frac{12}{\alpha^2 N^2}, \quad
n_s = 1 - \frac{9}{2 \alpha^2 N^2} - \frac{2}{N}.
\label{Indices}
\end{eqnarray}
%%%%%%%%%%%%%%%%%%%%%%%%%%%%%%%%%%%%%%%%%%%%%%%%%%%%%%%%%%%%%%%%%%%  
As a remark, in deriving this result, there appears one subtle point to be checked carefully.
Namely, we have approximated the factor $e^{- \alpha \sqrt{\frac{2}{3}} \varphi}$ by the number of
e-folding $N$ through the condition $\epsilon_V(\varphi_e) = 1$ as
%**   Non-e-varphi  %%%%%%%%%%%%%%%%%%%%%%%%%%%%%%%%%%%%%%%%%%%%%%%%%%%%%%%%%
\begin{eqnarray}
e^{\alpha \sqrt{\frac{2}{3}} \varphi} = \frac{(2n)^\alpha \alpha^2}{15} N 
+ \frac{(2n)^\alpha \alpha}{10 \sqrt{3}}
\approx \frac{(2n)^\alpha \alpha^2}{15} N.
\label{Non-e-varphi}
\end{eqnarray}
%%%%%%%%%%%%%%%%%%%%%%%%%%%%%%%%%%%%%%%%%%%%%%%%%%%%%%%%%%%%%%%%%%% 
The last approximation obviously depends on the size of $\alpha$ because of $0 < \alpha < 1$.
We will show the validity of this approximation shortly by taking numerical values concretely\footnote
{Of course, we can derive the similar result without using the approximation  (\ref{Non-e-varphi}).
Then, the expressions in Eq.  (\ref{Indices}) take more complicated forms  
%**   Indices2 %%%%%%%%%%%%%%%%%%%%%%%%%%%%%%%%%%%%%%%%%%%%%%%%%%%%%%%%%
\begin{eqnarray}
\epsilon_V &=& \frac{3}{4 \left(\alpha N + \frac{\sqrt{3}}{2} \right)^2}, \quad
\eta_V = - \frac{\alpha}{\alpha N + \frac{\sqrt{3}}{2}}, \nonumber\\
r &=& \frac{12}{\left(\alpha N + \frac{\sqrt{3}}{2} \right)^2}, \quad
n_s = 1 - \frac{9}{2 \left(\alpha N + \frac{\sqrt{3}}{2} \right)^2} - \frac{2 \alpha}{\alpha N + \frac{\sqrt{3}}{2}}.
\label{Indices2}
\end{eqnarray}
%%%%%%%%%%%%%%%%%%%%%%%%%%%%%%%%%%%%%%%%%%%%%%%%%%%%%%%%%%%%%%%%%%%
}.

In order to show that the result (\ref{Indices}) does not match the BICEP2 experiment,
let us impose the condition $r = 0.16$ from the BICEP2 results. Then, we have $\epsilon_V = 0.01$
owing to the definition $r = 16 \epsilon_V$. Using the definition of the spectral index,
$n_s \equiv 1 - 6 \epsilon_V + 2 \eta_V$ and the result $\eta_V = - \frac{1}{N}$ in Eq. (\ref{Indices}),
for $N = 60$ the spectral index is given by $n_s = 0.91$, which does not coincide with the BICEP2
result $n_s = 0.96$. Incidentally, using the result $r = \frac{12}{\alpha^2 N^2}$ in Eq. (\ref{Indices}), 
we can evaluate $\alpha = \frac{\sqrt{3}}{12}$, thereby making it possible to show 
$\frac{(2n)^\alpha \alpha^2}{15} N \gg \frac{(2n)^\alpha \alpha}{10 \sqrt{3}}$ as promised in the
above.

Alternatively, we can present the same conclusion by fixing the value of the spectral index $n_s$ first, and
then calculate the tensor-to-scalar ratio $r$. With the values $n_s = 0.96$ from the BICEP2 and
$N = 60$, the tensor-to-scalar ratio $r$ can be evaluated via Eq. (\ref{Indices}) to 
$r \simeq 0.027$, which is much smaller than the BICEP2 result $r = 0.16$.

At this stage, it is worthwhile to consider the role of the non-analytical functions in a general framework.
The point is that the non-analytical functions in general change the coefficient in the exponential
in the potential, that is, the exponential is changed from $e^{- \sqrt{\frac{2}{3}} \varphi}$ to 
$e^{- a \varphi}$ where $0 < a < \sqrt{\frac{2}{3}}$. Thus, in order to show that general
non-analytic functions do not reproduce the BICEP2 results, it is sufficient to consider a general
potential in a large field limit $\varphi \gg 1$
%**   G-Potential  %%%%%%%%%%%%%%%%%%%%%%%%%%%%%%%%%%%%%%%%%%%%%%%%%%%%%%%%%
\begin{eqnarray}
V(\varphi) =  V_0 \left( 1- c e^{- a \varphi} \right),
\label{G-Potential}
\end{eqnarray}
%%%%%%%%%%%%%%%%%%%%%%%%%%%%%%%%%%%%%%%%%%%%%%%%%%%%%%%%%%%%%%%%%%%   
where both $V_0$ and $c$ are positive. It is easy to see that this more general potential
produces the same result as (\ref{Indices}) in the more specific potential (\ref{Non-V}).

%%%%%%%%%%%%%%%%%%%%%%%%%%%%%%%%%%%%%%%%%%%%%%%%%%%%%%%%%%%%%%%%%%%%%
%%%%%%%%%%%%%%%%%%%%%%%%%%%%%%   SEC  7    %%%%%%%%%%%%%%%%%%%%%%%%%%
%%%%%%%%%%%%%%%%%%%%%%%%%%%%%%%%%%%%%%%%%%%%%%%%%%%%%%%%%%%%%%%%%%%%%
\section{Conclusion}

In this article, we have investigated a locally scale-invariant theory which describes the coupling of gravity 
and the standard model in detail. After constructing an abelian gauge model, it is shown that this model
exhibits a peculiar feature of a coupling between the gauge field and the inflaton in large field
values while it describes the standard model coupled to general relativity in small field values.
Moreover, we have extended the abelian model to non-Abelian gauge groups. It is also straightforward to
generalize our model to the case of multi-component scalar fields.

We have also discussed a possibility that our model could explain the recent BICEP2 results, but
we have shown that it is very difficult to do so even if we take into consideration non-analytic types 
of the potential. This fact is naturally understood since our model shares many of physical properties 
with the Starobinsky $R^2$ inflation model \cite{Starobinsky}.
 
One interesting direction of future work is to spell out effects on the reheating coming from the coupling
between the gauge field and the inflaton \cite{Tomoyose}. Another future problem is to clarify quantum effects 
in our model along our previous study \cite{Oda-Tomo}. The latter problem is very important since
we have to show that the running coupling constant $\lambda(\mu)$ could take the value $\lambda = {\cal O}(0.1)$ at the TeV
scale to make our model to be consistent with the standard model. 
We wish to return these problems in near future.

%%%%%%%%%%%%%%%%%%%%%%%%%%%%%%%%%%%%%%%%%%%%%%%%%%%%%%%%%%%%%%%%%%
%%%%%%%%%%%%%%%%%%%%%%%% Acknowledgements %%%%%%%%%%%%%%%%%%%%%%%%%%%%%
%%%%%%%%%%%%%%%%%%%%%%%%%%%%%%%%%%%%%%%%%%%%%%%%%%%%%%%%%%%%%%%%%%
\begin{flushleft}
{\bf Acknowledgements}
\end{flushleft}
This work is supported in part by the Grant-in-Aid for Scientific 
Research (C) No. 25400262 from the Japan Ministry of Education, Culture, 
Sports, Science and Technology.

%%%%%%%%%%%%%%%%%%%%%%% reference %%%%%%%%%%%%%%%%%%%%%%%%%%%%%%%
%%%%%%%%%%%%%%%%%%%%%%%%%%%%%%%%%%%%%%%%%%%%%%%%%%%%%%%%%%%%%%%%%%

\end{document}